\documentclass[10pt,twocolumn]{IEEEtran}
\usepackage{amsmath}
\usepackage{amsthm,amssymb,amsmath,bm}
\usepackage{subfigure}
\usepackage{amsfonts}
\usepackage{epsfig}
\usepackage{amssymb}
\usepackage{amsmath}
\usepackage[table,xcdraw]{xcolor}
\usepackage{cite}
\usepackage[utf8x]{inputenc}
\usepackage{color,soul}
\usepackage{subfigure}
\usepackage{multirow}
\usepackage{rotating}
\usepackage{graphicx}
\usepackage{tabularx}
\usepackage{array}
\usepackage{color,soul}
\usepackage{bm}
\usepackage{graphicx,dblfloatfix}
\usepackage{blindtext}

\usepackage{subfigure}
\usepackage{moreverb}
\usepackage{epsfig}
\usepackage{amsmath,amssymb,amsthm,mathrsfs,amsfonts,dsfont}
\usepackage{adjustbox,lipsum}
\usepackage{amsfonts}
\usepackage{epsfig}
\usepackage{amssymb}
\usepackage{amsmath}
\usepackage{amsthm}
\usepackage{subfigure}
\usepackage{multirow}
\usepackage{rotating}
\usepackage{graphicx}
\usepackage{tabularx}
\usepackage{array}
\usepackage{anyfontsize}
\usepackage{color,soul}
\usepackage{graphicx,dblfloatfix}
\usepackage{epstopdf}
\usepackage{blindtext}
\usepackage{amsmath}
\usepackage{amsthm,amssymb,amsmath,bm}
\usepackage{subfigure}
\usepackage{amsfonts}
\usepackage{epsfig}
\usepackage{amssymb}
\usepackage{amsfonts}
\usepackage{amsmath}
\usepackage{cite}
\usepackage{graphicx}
\usepackage{fancyhdr}
\usepackage{subfigure}
\usepackage[subfigure]{tocloft}
\usepackage[font={small}]{caption}
\usepackage{subfigure}
\usepackage{tabularx}
\usepackage{tcolorbox}
\usepackage{cite}
\usepackage[linesnumbered,ruled,vlined]{algorithm2e}
\SetKwInput{KwInput}{Input}
\SetKwInput{KwOutput}{Output}
\usepackage{amsthm,amssymb,amsmath,bm}
\usepackage{graphicx}
\usepackage{fancyhdr}
\usepackage{subfigure}
\usepackage[subfigure]{tocloft}
\usepackage[font={small}]{caption}
\usepackage{subfigure}
\usepackage{tabularx}
\usepackage{cite}

\usepackage{stackengine}

\allowdisplaybreaks
\usepackage[colorlinks,bookmarksopen,bookmarksnumbered,citecolor=red,urlcolor=red]{hyperref}

\begin{document}
	\title{ \huge AI-based Radio Resource Management and Trajectory Design for PD-NOMA Communication in IRS-UAV Assisted Networks}	
	\author{\IEEEauthorblockN{Hussein M. Hariz, Saeed Sheikhzadeh, Nader Mokari, 
			Mohammad R. Javan, 
			B. Abbasi-Arand and Eduard A. Jorswieck }
		\thanks{Hussein M. Hariz, Saeed Sheikhzadeh, Nader Mokari and B. Abbasi-Arand are with the Department of Electrical and Computer Engineering, Tarbiat Modares University, Tehran, Iran (e-mail: nader.mokari@modares.ac.ir, abbasi@modares.ac.ir).
			Mohammad~R.~Javan is with Department of Electrical and Computer Engineering, Shahrood University of Technology, Iran (e-mail: javan@shahroodut.ac.ir).
			Eduard A. Jorswieck is with the Department of Information Theory and Communication Systems, Technische Universit\" at Braunschweig, Braunschweig, Germany (e-mail: jorswieck@ifn.ing.tu-bs.de). This work was
			supported by the joint Iran national science foundation (INSF) and German research foundation (DFG) under grant No. 96007867.
	} }
	
	\maketitle
	
	\vspace{-2cm}
	\begin{abstract}
		In this paper, we consider that the unmanned aerial vehicles (UAVs) with attached intelligent reflecting surfaces (IRSs) play the role of flying reflectors that reflect the signal of users to the destination, and utilize the power-domain non-orthogonal multiple access (PD-NOMA) scheme in the uplink. We investigate the benefits of the UAV-IRS on the internet of things (IoT) networks that improve the freshness of collected data of the IoT devices via optimizing power, sub-carrier, and trajectory variables, as well as, the phase shift matrix elements. We consider minimizing the average age-of-information (AAoI) of users subject to the maximum transmit power limitations, PD-NOMA-related restriction, and the constraints related to UAV’s movement. The optimization problem consists of discrete and continuous variables. Hence, we divide the resource allocation problem into two sub-problems and use two different reinforcement learning (RL) based algorithms to solve them, namely the double deep Q-network (DDQN) and a proximal policy optimization (PPO). Our numerical results illustrate the performance gains that can be achieved for IRS enabled UAV communication systems. Moreover, we compare our deep RL (DRL) based algorithm with matching algorithm and random trajectory, showing the combination of DDQN and PPO algorithm proposed in this paper performs 10\% and 15\% better than matching algorithm and random-trajectory algorithm, respectively.
	\end{abstract}
	\section{Introduction}
	\subsection{State of the art}
	Unmanned aerial vehicles (UAVs) play an important role in enhancing spectral efficiency \cite{gupta2015survey}. The UAVs can improve the telecommunication system from different aspects, e.g., they can improve the coverage area and rate  \cite{ effic_uavs }, increase secure communications\cite{secur_uavs}, achieve low cost and high mobility\cite{ effect_energy}, achieve long-range control in real time \cite{secur_uavs}, they can also decrease outage probability \cite{ wireless_uavs},
	or provide on-demand deployment \cite{wireless_uav}. 
	Another major application is to employ UAVs as mobile relays to provide reliable connectivity for remote users in locations where there are no established wireless networks (e.g., an empty desert) \cite{throughput_uav} and \cite{UAV_relay}.  
	
	Recently, power-domain non-orthogonal multiple access (PD-NOMA) has been considered as a promising solution to improve the communication throughput \cite{PD-NOMA}. The basic idea behind the PD-NOMA technique  is that more than one user can simultaneously use a sub-carrier. The combination of the PD-NOMA and UAVs  was considered as a solution to improve the coverage and spectral efficiency (SE) \cite{joint-NOMA}.  Moreover, the intelligent reflecting surface (IRS), known as a large intelligent surface (LIS),  is an artificial radio structure which reflects the incident radio-frequency (RF) waves into specified directions \cite{tan2018enabling} and \cite{lmc2019intelligent}. These surfaces can be manufactured at appropriate cost and as such they can be deployed universally providing an unprecedented opportunity to control the wireless multi-path radio environment. 
	The combination of UAV and IRS is another way of improving the efficiency of UAV application \cite{IRS-UAVs}.

	In order to characterize the freshness of information, the concept of age of information (AoI) has been introduced. The AoI is defined as the amount of time that has elapsed since the source generated the last successfully received update packet at the monitor \cite{AOI_Real}. In other words, the AoI is a metric to quantify the freshness of information at the destination. 
	
	
	With the growth of the industrial field, the industrial internet of things (IIoT) plays an important role in connecting the physical objects in the industry environment with the internet \cite{IIoT}.
	We can also benefit the UAVs in the IIoT areas 
	\cite{priority_uav}. 
	The combination of UAV and IIoT seems to be essential for the future industry, due to the highly changing environment, as well as, need for immediate response and time-restricted data. 

	
	
	\subsection{Related works}
	
	In the following part, with consider the UAV as an effective technique to improve the performance of the network, firstly, we investigate the applications of UAV in the IoT and IIoT area. Then, we summarize the works that existed in the UAV and AoI sectors. After that, the combination of the IoT and AoI with the presence of the UAVs is considered. Finally, the applications of artificial intelligence in those areas are argued.
	\begin{itemize}
		\item \textbf{UAV, IoT, and IIoT:}
		The authors in \cite{DSF-NOMA} utilize UAVs to assist with emergency communication in a heterogeneous IoT (Het-IoT) and considered the NOMA scheme without the necessity of successive interference cancellation (SIC). The authors in\cite{3D-IOT} study the three-dimensional (3D) placement and resource allocation of multiple UAV-mounted base stations (BSs) in an uplink IoT network and proposed a clustering algorithm to achieve high reliability, and effectiveness.
		\item \textbf{UAV and AoI:}
		The wireless network assisted by the AoI-aware UAVs has received increasing attention, due to the importance of the AoI and the advantages of the UAVs in IoT systems \cite{liu2018age,jia2019age,abd2018average}.
		In \cite{liu2018age}, the authors proposed the UAV’s trajectory  to keep the sensed data fresh in wireless sensor networks. This paper optimizes the minimization of the average AoI (AAoI) of various wireless sensors, where the UAV is used to collect the ground sensor nodes (SNs) data.
		The authors in \cite{jia2019age} proposed  joint optimization of the UAV trajectory and data collection  based on dynamic programming. The paper try to minimize AoI of the SNs, independent form the UAV flight mode. 
		The authors in \cite{abd2018average} proposed an efficient iterative algorithm to minimize the total AoI of the system for the UAV’s flight trajectory and service time allocation jointly.
		\item \textbf{UAV, IoT, and AoI:}
		The authors in \cite{AoI_UAV_iot} investigate the UAV-assisted wireless powered IoT system and study a joint energy transfer and data collection time allocation and a UAV’s trajectory planning problem.
		In \cite{abd2018average}, the authors investigate the role of a UAV as a mobile relay to minimize the average peak age-of-information (PAoI) for a source-destination pair and jointly optimize the UAV’s flight trajectory as well as energy and service time allocations for packet transmissions.
		\item \textbf{UAV, IoT, and AI:}
		In \cite{MEC-IOT}, the authors proposed a cooperative UAV-enabled mobile edge (MEC) network structure in which the UAVs are able to support other UAVs to perform the computation tasks, while considering the interference reduction from UAVs to devices, where the deep reinforcement learning (DRL) based algorithms to achieve better performance than that of the non-cooperative UAV edge computing methods. The authors in \cite{Learning-IIoT} study an UAV-enabled system architecture consisting of multiple IIoTs, in which the collected data by sensors in IIoTs can be delivered to UAVs for processing directly and also  proposed  a learning-based cooperative particle swarm optimization algorithm (LCPSO) with Markov random field (MRF) based decomposition strategy.
		\item \textbf{UAV, AoI, and AI:}
		The authors in \cite{UAV_AOI_AI} considered UAV-assisted single-hop vehicular network, DRL and leverage Deep Deterministic Policy Gradient (DDPG) to learn UAVs trajectory and scheduling policy under minimum throughput constraints, and also the deployed UAVs adapt their velocities during the data collection mission in order to minimize the AoI.
	\end{itemize}
	
	We summarize the main differences between our paper and other related works in Table \ref{tab:Related_work}.
	\begin{table*}[!h]
		\centering
		\caption{ Related work and contribution}
		\label{tab:Related_work}
		
		\begin{tabular}{|p{0.5cm}|m{1cm}|m{1.5cm}|p{1cm}|p{2cm}|p{2.5cm}|p{3.1cm}|p{2.5cm}|}
			\hline
			\textbf{Ref.} & \textbf{Location of IRS} & \textbf{Access technology} &\textbf{Number of UAV}  & \textbf{Duty of UAV} & \textbf{Objective function} & \textbf{Constraints} & \textbf{Machine learning based solution} \\
			\hline
			\cite{liu2018age}	& No IRS & OMA  & One  &  Collecting and transmitting. &  Minimize the maximum AoI and Minimize the AAoI.& Trajectory  &  DP and GA.\\ 
			\hline
			\cite{AoI_UAV_iot}	& No IRS & OMA  &One   &Collecting and transmitting.  & Minimize the AAoI.  &  Trajectory, allocate ‎the time for EH and ‎data collection for all ‎SNs. & KKT, DP, and	AC.‎
			\\ 
			\hline 
			\cite{PPO-RIS-AoI}& On UAV. & OMA  & One  & Relaying & Minimize  the EAoI.‎
			&UAV altitude and the ‎IoTDs scheduling.  & DRL, MDP, and PPO.   \\
			\hline
			\cite{UAV_AOI_AI}	& No IRS & TDMA & Multiple-‎UAVs. &Collecting and processing.
			& Minimize the EWSA.‎  &  Trajectories and scheduling ‎policies. & DRL and DDPG.‎ \\
			\hline
			\cite{al2021ris}	& On fixed point. & FDMA  & One & Collecting and transmitting. & Maximize the total number of served devices. ‎ & UAV trajectory, ‎IoTDs scheduling and RIS phase-shift. & MDP, PPO, and BCD. \\
			\hline
			\cite{pan2021uav} & On fixed point.  & OMA  &  One & UAV-‎supported ‎THz ‎communication .& Maximize the minimum average achievable rate.‎ & UAV’s trajectory, the	phase shift of IRS, the ‎allocation of THz ‎band and the power ‎control.
			& No.‎ \\
			\hline
			\cite{AoI-ARIS-1} & No IRS &  TDMA & One  &Collecting and transmitting.   & Minimize the maximum and AAoIs. &  Trajectory and design of ‎SN association. & APC, DP, and GA. ‎
			\\
			\hline
			\cite{wany_packet}	&No IRS& OMA & One  &Collecting and transmitting. & Minimize the total number of expired packets and	minimize the maximum AoI.
			& Trajectory and expired packets.  & RL. \\
			\hline
			\cite{Zhang_AOI_sensing}&No IRS&OMA&One&Collecting and transmitting. & Minimize the AoI.
			&  Sensing time,‎ transmission time, UAV ‎trajectory and task ‎scheduling optimization problem.‎
			&RL and DP.‎
			\\ 
			\hline
			\cite{Cooperative}&No IRS& OMA & Multiple-‎UAVs.& Collecting and transmitting & Minimize the AoI. & The UAV trajectory design. & CA2C, DDPG, and DQN.  \\ 
			\hline
			\cite{Wu1}	& No IRS& OMA & Multiple-‎UAVs. &  Collecting and transmitting. & Minimize the AAoI. & Trajectory of the UAV in the sensing stage and transmission stage. &  DRL and DDQN.\\
			\hline
			\cite{Renzo_UAV}& On fixed point &  OMA & One  & Improve the throughput of wireless networks. & Improve the average achievable rate	of the system. & Trajectory design and Passive beamforming. & No\\ 
			\hline
			Our Paper & On UAV & PD-NOMA  & Multiple-UAVs. & Relaying & Minimizing the AAoI.  & Trajectory, reflection amplitude and phase shift, transmit power of users and sub-carrier assignment. &Combination of PPO and DDQN methods. \\ 
			\hline
		\end{tabular}
	\end{table*}
	\subsection{Contribution and motivation}
	
	The most related works to our paper are \cite{PPO-RIS-AoI}, \cite{muhammad2021age}, and \cite{UAV-AoI-IoT-sch}. In \cite{PPO-RIS-AoI}, the authors determined the scheduling and altitude. However, the trajectory is not considered, where we consider the sub-carrier allocation and trajectory. The authors considered just one UAV, but we use several UAVs to improve the converge of users. Moreover, the authors in \cite{PPO-RIS-AoI} neglect the direct channel from the BS to users and use orthogonal multiple access (OMA) techniques. However, besides the non-direct link, we consider a direct link between users and the receiver, as well as, the PD-NOMA technique.
	In the solution part, the authors use the RL method to solve the proposed problem. Furthermore, we propose combinatorial machine learning methods where discrete and continuous actions are derived separately.
	
	The authors in \cite{muhammad2021age} considered the joint IRS phase shift and scheduling, but there is no UAV and the location of IRS is fixed. In comparison, we use the IRS attached UAVs to optimize the channels. The IRS-UAVs can fly vertically and find better situations to improve the performance of the network. 
	Also, in \cite{muhammad2021age}, the authors proposed algorithm based on semi-definite relaxation (SDR) to solve the problem. As mentioned before, we adopt two mixed DRLs based methods. 
	
	In \cite{UAV-AoI-IoT-sch}, the authors considered joint scheduling policy and dynamic UAV altitude control in the UAV-assisted IoT networks without considering an IRS.
	It is worth mentioning that the combination of UAV and IRS is a good method to improve the coverage and capacity of the mobile networks that we consider in our paper.
	The authors in \cite{UAV-AoI-IoT-sch} neglect the direct channel between the IoT and BS, and 
	also the OMA technique is adopted. 

	The main contributions of our paper can be summarized as follows points: 
	\begin{itemize}
		\item{We propose a novel model in the IRS which is embedded in the UAVs. The IRS-UAV assisted networks improve the uplink channel reliability between the IoT devices and a central unit (CU) as a receiver and enhances the freshness of information. 
			Also, we propose the AAoI formulation for the IRS-UAV assisted networks under different practical constraints. } 
		\item {We formulate the optimization problem to find the optimal variables such as trajectory, the configuration of IRS (reflection amplitude and phase shift), power allocation, and sub-carrier assignment. The objective function of our problem is minimizing the AAoI of users over time slots subject to the maximum transmit power, PD-NOMA-related constraints, and restrictions related to UAVs trajectory.}
		\item {We formulate the status update framework for IoT users in an aerial IRS problem as a Markov Decision Process (MDP) and adopt the DRL based learning method to learn environment dynamics in order to handle the trajectory of the UAVs, power allocation, and sub-carrier assignment. The proposed problem consists of discrete and continuous variables,  
			for that we leverage the DDQN for discrete actions and the PPO algorithm for continuous actions. This solution is a highly stable state-of-the-art model-free DRL and it can find the best policy that minimizes the AAoI.}
		\item{Numerical results reveal significant improvement by comparing with other algorithms and a benchmark method. The performance gains achieved by our proposed designs provide insights into the practical deployment of IRS-UAV-assisted IoT-enabled networks.}
	\end{itemize}

	\textit{Notation:} Matrices are shown by bold uppercase letter, such as $\boldsymbol{H}$ and vectors are denoted by lowercase letters, such as $\boldsymbol{h}$. $\mathcal{C}^{M\times N}$ and $\mathcal{C}^{M\times 1}$ represent the set of all  $M\times N$ matrices with complex elements and the set of all $M\times 1$ vectors with complex arrays, respectively.   $(\cdot)^T$ and $(\cdot)^H$ show the transpose and conjugate transpose, respectively. The notation $\text{diag}(\boldsymbol{h})$ represents a diagonal matrix with elements in vector $\boldsymbol{h}$. $\mathbb{E}$ and $\parallel\cdot\parallel$ denote the expectation  and Euclidean norm operations, respectively. $\mathcal{CN}(\mu,\sigma^2)$ shows the complex circular Gaussian random variable with mean $\mu$ and variance $\sigma^2$.  We show the abbreviations in Table \ref{abbreviations}.

	\begin{table}
		\caption{List of abbreviations in alphabetical order.‎}
		\label{abbreviations}
		\centering
		\begin{tabular}{|p{1.5cm}|p{6cm}|}
			\hline
			\textbf{Item} & \textbf{Description}  \\
			\hline
			AAoI& Avarge  Age of Information.\\‎	
			\hline 
			AI &Artificial Intelligence.‎\\	
			\hline 
			AoI& Age of Information.‎\\	
			\hline 
			AP & Access Point. ‎   \\
			\hline
			APC&Affinity ‎Propagation Clustering.\\
			\hline 	
			BCD &Block Coordinate ‎Descent.‎	\\‎ 
			\hline 
			CA2C&  Compounded-Action Actor-Critic.‎	\\
			\hline
			CU& Central Unit.‎	\\
			\hline 
			DDPG& Deep Deterministic Policy Gradient.‎	\\
			\hline 
			DDQN& Double Deep Q-Network. ‎	\\
			\hline 
			DNNs &Deep Neural Networks.‎	\\
			\hline
			DP&Dynamic Programming.‎	\\
			\hline  
			DQN &Deep Q-Network.‎	\\
			\hline  
			DRL& Deep Reinforcement Learning.‎	\\
			\hline 
			EAoI& Expected Sum AoI.‎\\	
			\hline
			EH &Energy Harvesting.‎	\\
			\hline 
			
			EWSA& The Expected Weighted Sum AoI.‎	\\
			\hline 
			GA‎&  Genetic ‎Algorithm.‎	\\
			\hline 
			IIoT & Industrial Internet of Things.‎	\\
			\hline 
			IRS& Intelligent Reflecting Surface.‎	\\
			\hline 
			IoT& Internet of Things.‎	\\
			\hline 
			KKT& Karush-Kuhn-Tucker.‎	‎ \\
			\hline 
			LCPSO& Learning Based Cooperative Particle ‎Swarm Optimization Algorithm.‎	\\
			\hline
			LIS &Large Intelligent Surface.‎	\\
			\hline 
			MDP& Markov Decision Process.‎	\\
			\hline
			MEC & Mobile Edge Computing. ‎	\\
			\hline  
			MRF & Markov Random Field. ‎	\\
			\hline 
			NLoS & Link  is Non-Line of Sight.‎	\\
			\hline 
			NNs & Neural Networks.‎	\\
			\hline 
			NOMA & Non Orthogonal Multiple Access.‎\\	
			\hline 
			OMA& Orthogonal Multiple Access.‎	\\
			\hline 
			PD-NOMA& Power Domain Non Orthogonal Multiple ‎Access.‎\\	
			\hline 
			PPO &Proximal Policy Optimization.‎	\\
			\hline 
			RF &Radio-Frequency.‎	\\
			\hline 
			RL &Reinforcement Learning .‎	\\
			\hline 
			SDR &Semi-Definite Relaxation.‎	\\
			\hline 
			SE&Spectral Efficiency. ‎	\\
			\hline
			SIMO&Single Input Multiple Output.‎	\\
			\hline
			SINR &Signal To Noise Plus Interference Ration. ‎\\	
			\hline 
			SIC & Successive Interference ‎Cancellation. ‎	\\
			\hline 
			UAV& Unmanned Aerial Vehicle.‎	\\
			\hline 
			UE &User Equipment.‎	\\
			\hline                  	
		\end{tabular}
	\end{table}

	\subsection{{Paper Organization}}
	This paper is arranged as follows. In Section \ref{SM_PF}, the system model and problem formulation are proposed. Section \ref{Solution} presents the solution method of the formulating problem. Simulation results are provided in Section \ref{Simulation_Results}. In the end, concluding remarks of this paper are stated in Section \ref{CONCLUSION}.

	\section{System Model and Problem Formulation}\label{SM_PF}
	In this section, the considered network, related parameters, and channel models of uplink PD-NOMA IRS-based UAV communication system are discussed. Then, the resource allocation problem 
	is introduced.

	\subsection{System Model and Parameters}
	
	\begin{figure}[t]
		\centering
		{\includegraphics[width=0.5\textwidth]{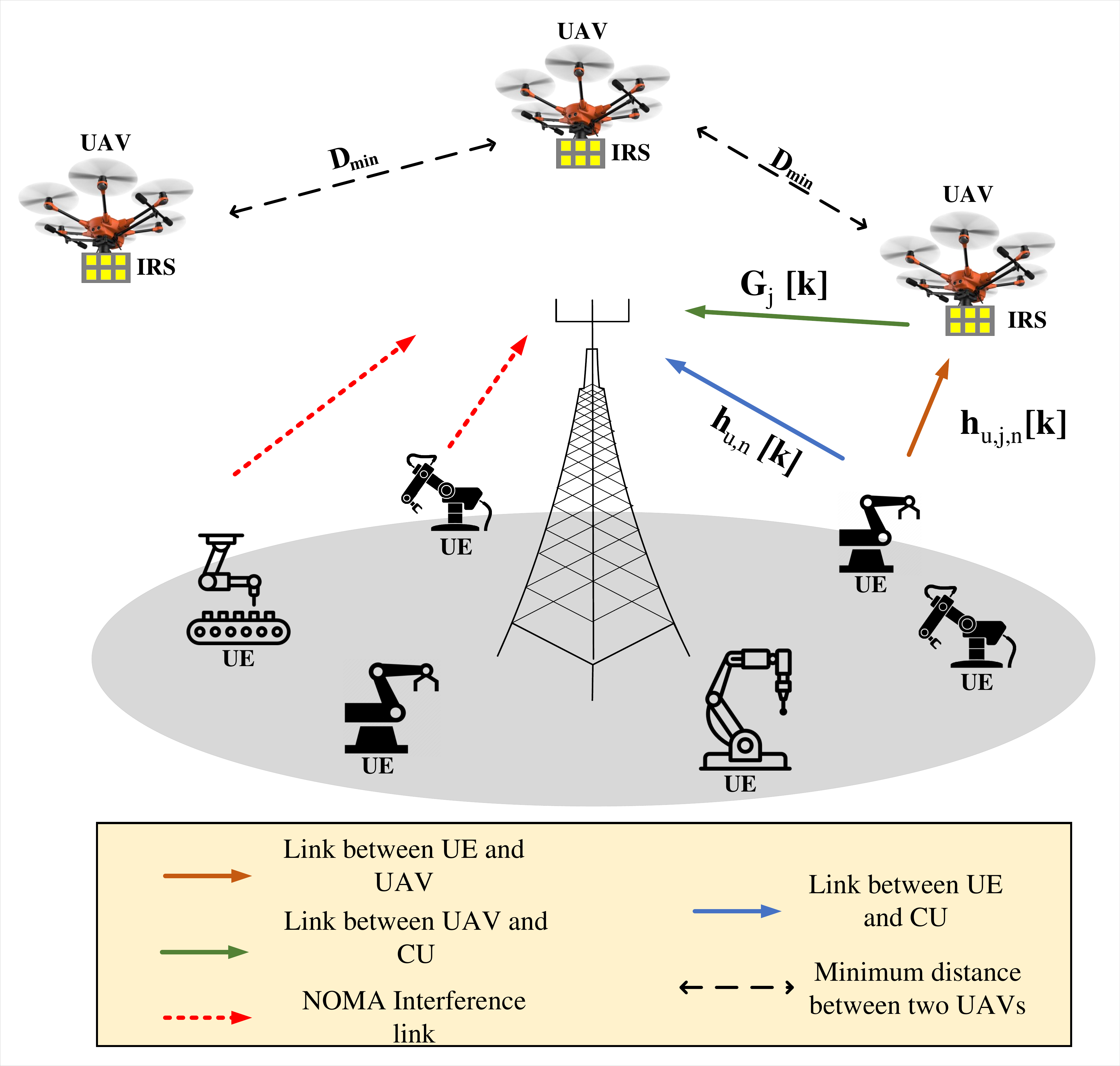}}
		\caption{System model consisting of $J$ UAVs equipped with $L$ IRS's components and help IoT devices to transmit their own data toward the CU. }
		\label{System_Model}
	\end{figure}

	We consider the uplink transmission of IRS-assisted UAVs communication system to serve IoT user equipments (UEs) on the ground. Each UE sends the measured data to a central unit (CU) through SIMO network.
	
	As shown in Fig. \ref{System_Model}, we consider $U$ single antenna of UEs are denoted by set $\mathcal{U}=\{1,\ldots ,U\}$, and indexed by $u\in\mathcal{U}$. The location of UE $u$ is denoted as $\boldsymbol{\alpha}_u=[x_u,y_u,0]^{T}$. The users are stable in their location. 
	The total bandwidth is divided into $N$  orthogonal sub-carriers denoted by $ n \in \mathcal{N}=\{1,\cdots,N\} $.
	
	It is assumed that there is a single CU as a macro base station (MBS), with $M$ antennas indexed by $m$ and 
	locates at the origin. In our model, all UEs send their own signals to the CU directly and/or via reflection from UAVs. 
	We suppose that there are $J$ flying UAVs in the network. The set of UAVs is indicated as $\mathcal{J}=\{1,\dots,J\} $ and each one is indexed by $j$.

	Each UAV is equipped with IRS consisting of $L$ reflecting elements with an equal distance $d$(\textrm{mm}). It is worth mentioning that these elements are active. Hence, they can change the value of the reflecting amplitude and phase shift.
	Here, the UAV-IRS plays the role of a flying reflector that reflects the signal of UEs to the  CU. 
	Let $\boldsymbol{\Theta}_j=\operatorname{diag}\left(\beta_{1,j} e^{i \theta_{1,j}}, \cdots, \beta_{L,j} e^{i \theta_{L,j}}\right)$ denotes the reflection-coefficient matrix in UAV $j$, where $\beta_{l,j} \in[0,1]$ and $\theta_{l,j} \in[0,2 \pi]$ denote the reflection amplitude and phase shift of  $l$-th element, respectively \cite{Ru_Weighted}. Here, we consider both of them as the optimization variables. 
	
	We consider that the time is divided into $K$ discrete time slots with equal duration $\delta$. The location of flying UAV $j$ at time slot $k$ can be denoted as $\boldsymbol{q}_j[k]=[x_j[k], y_j[k],z_j]^T$. We assume that the location of UAVs changes with respect to the X-axis and Y-axis and is fixed related to Z-axis. 
	The UAV trajectory at each time slot needs to meet the following constraints:
	\begin{align}
		\label{UAV_Con}\text{C1:~}& \|\boldsymbol{q}_j[k]-\boldsymbol{q}_j[k-1]\|^{2} \leq D^{}, ~~~\forall k, \forall j\in\mathcal{J},\\
		\label{UAV_Con_2}\text{C2:~}&\|\boldsymbol{q}_j[k]-\boldsymbol{q}_{j'}[k]\|^{2} \geq D_{\min}, ~~~~\forall k, \forall j',j\in\mathcal{J},\\
		\label{UAV_Con_3}\text{C3:~}&\|\boldsymbol{q}_j[k]\|^2\leq r_{\max}, ~~~~~~~~~~~~~~~\forall k, \forall j\in\mathcal{J},
	\end{align}
	where $D_{\min}$ is the minimum distance between any pair of UAVs, $ r_{\max} $ is the maximum radius of coverage CU, and $D = v_{\max} \delta$ is the distance that UAV $j$ with maximum speed $v_{\max}$ can move during each time slot. \eqref{UAV_Con} is related to the maximum distance that UAV $j$ during time slot $k$ can horizontally move, \eqref{UAV_Con_2} ensures UAVs never have collision, and \eqref{UAV_Con_3} guarantees that the UAV stays in the coverage area of CU \cite{Renzo_UAV}.   
	
	
	In the following, we will discuss the broadband channel in the link between the UEs, UAVs, and CU. Here, we have three kinds of channels: 1) channel between the CU and UEs, 2) channel between the UEs and UAV, and 3) channel between the UAV and CU. 
	The channel coefficient between UE $u$ and the CU on sub-carrier $n$ in time slot $k $ is denoted by $\bold{h}_{u,n,(BS)}[k]\in\mathcal{C}^{M\times 1 }$. We assume that this link is non-line of sight (NLoS) and experiences an independently distributed Rayleigh fading. The channel gain can be modeled as follows \cite{Renzo_UAV}:
	\begin{align}
		\bold{h}_{u,n,(BS)}[k]=\sqrt{\rho d_u^{-\alpha}}\boldsymbol{g}_{u,n}[k],
	\end{align}
	where $ \rho$ is the path loss at the reference distance $ D_0=1$m, $ \alpha $ is the path loss exponent, and $d_u$ indicates the distance between the CU and UE $ u $ at slot $ k $. $\boldsymbol{g}_{u,n}[k]$ is the small scale fading and is generated with complex Gaussian distribution with zero mean and unit variance, and the location of the CU and UEs are fixed over time slots.

	The channel gain between UAV $j$ and UE $ u $ over sub-carrier $n$ at time slot $ k $ is denoted by
	$ \bold{h}_{u,j,n}[k]\in\mathcal{C}^{L\times 1} $. The UAVs fly at the higher altitude of $z_u$ meters from the ground level, hence, we consider that there is LoS-dominated propagation in the link between the UEs and UAVs, as well as the UAVs and CU. As described before, each UE  is in a specific position $ \boldsymbol{\alpha}_u $, far from the CU. Therefore, the time-dependent distance between  UE $u$ and  UAV $j$ denoted by $ d_{u,j}[k] $ and can be calculated by 
	\begin{align}
		d_{u,j}[k]={\|\boldsymbol{\alpha}_u-\boldsymbol{q}_j[k]\|}. 
	\end{align} 
	The channel gain includes path loss, small-scale fading, and IRS array response and can be given by \cite{Renzo_UAV}
	\begin{align}\nonumber
		&\bold{h}_{u,j,n}[k]
		=
		\\&\underbrace{\sqrt{\rho d_{u,j}[k]^{-\alpha}}}_{\text {path loss }}\underbrace{\left(\sqrt{\frac{\kappa_n}{1+\kappa_n}} \bold{h}_{u,j}^{\mathrm{LoS}}[k]+\sqrt{\frac{1}{\kappa_n+1}} \bold{h}_{u,j}^{\mathrm{NLoS}}[k]\right)}_{\mathrm{\text{aray response \& small-scale fading} }},
	\end{align}
	where $ \kappa_n$ is the Rician factor. $\bold{h}_{u,j}^{\mathrm{NLoS}}[k] $ is NLOS term and can be complex Gaussian distribution with  zero mean and unit variance. $ \bold{h}_{u,j}^{\mathrm{LoS}}[k]$ is the LoS term and is given by
	\begin{align}
		\bold{h}_{u,j}^{\mathrm{LoS}}[k]=\left[1, e^{-i \frac{2 \pi}{\lambda} d \psi_{u,j}[k]}, \ldots, e^{-i \frac{2 \pi}{\lambda}(L-1) d \psi_{u,j}[k]}\right]^{T} \in \mathcal{C}^{L \times 1},
	\end{align}
	where  $ d $ is the distance between antennas arrays, $ \lambda $ is wavelength, and $ \psi_{u,j}[k]=\frac{x_u-x_j[k] }{d_{u,j}[k]}$ is the cosine of the
	angle of departure \cite{Renzo_UAV}. 
	
	The signals of UAVs can propagate in obstacle-free airspace. In other words, the link between the UAVs and CU is existing, hence, the effect of reflection or obstruction can be ignored \cite{wei1}. The channel gain between the CU and IRS $l$  of UAV $ j $ at time slot $k$ is given by;  
	\begin{align}
		g_{m,j_l}[k]={\sqrt{\rho d_{\text{BS},j}[k]^{-\alpha}}} \exp \left(-i \frac{2 \pi}{\lambda} (l-1) d \psi_{\text{BS},j}[k]\right),
	\end{align}
	where $d_{\text{BS},j}[k]={\|\bold{q}_j[k]\|} $ and $\psi_{\text{BS},j}[k]=\frac{x_j[k]}{d_{\text{BS},j}[k]}$.  Let's denote $ \bold{G}_j[k] \in \mathcal{C}^{M\times L}, j \in \mathcal{J}$  as the channel gain matrix between  UAV $j$ and the CU with elements $ [g_{m,j_l}[k]]$.
	
	Now, the effective channel between UE $u$ and CU  assisted by UAV $j$ over sub-carrier $n$ denoted by $  \bold{h}_{(u,j,n),\text{BS}}^{H}[k]$ is equal to $ \bold{h}_{u,j,n}^{H}[k]\boldsymbol{\Theta}_j[k]\bold{G}_j[k] +\bold{h}_{u,n,(BS)}[k]$. Finally, the channel coefficient between each UE $u$ on sub-carrier $n$ through all the UAVs can be formulated by 
	\begin{align}
		\bold{h}_{u,n}[k]=\sum_{j=1}^{J}	\bold{h}_{u,j,n}^{H}[k]\boldsymbol{\Theta}_j[k]\bold{G}_j[k] +\bold{h}_{u,n,(BS)}[k].
	\end{align} 
	\\\indent 
	The sub-carriers are assigned to the UEs by utilizing the PD-NOMA schemes. Generally, in the PD-NOMA technique, each sub-carrier can be assigned to more than one UE in the service coverage area. Let us define sub-carrier assignment variable by $ \rho_{u}^{n}[k] \in\{0,1\}$. If sub-carrier $n$ is assigned to UE $u$ at time slot $k$, we denote  $\rho_{u}^{n}[k]=1$, otherwise, its value equals $ 0 $.  
	Based on the PD-NOMA technique, each sub-carrier can be utilized at most $Q$ times in the network. In other words, $Q$ is the cluster size of PD-NOMA and is guaranteed by the following:
	\begin{align}
		\text{C4:~}&\sum_{u=1}^{U}	\rho_{u}^{n}[k]\le Q,~~~~ \forall k,\forall n\in\mathcal{N}.
	\end{align}
	
	The power allocation variable of user $u$ on sub-carrier $n$  at slot $k$ is $ p_{u}^{n}[k]$. Here, we consider that the transmit power related to each user over the assigned sub-carrier has to remain in the predefined limitation. Moreover, there is a restriction related to the maximum power that each user can transmit. 
	~The power allocation constraints can be formulated as
	\begin{align}
		\label{k}
		\text{C5:~}&p_{u}^{n}[k]\rho_{u}^{n}[k] \le P_{\text{mask}}^{n}, ~~~~~~~~\forall n \in \mathcal{N}, \forall u \in\mathcal{U}, \forall k,\\
		\text{C6:~}&\sum\limits_{n=1}^{N} p_{u}^{n}[k]\rho_{u}^{n}[k] \le P^u_{\max} ,~~~\forall u \in\mathcal{U},\forall k,
	\end{align}
	where $P_{\text{mask}}^{n}$ and $P^u_{\max}$ are the maximum allowable transmit power of user $u$ over  sub-carrier $n$ and total transmit power of user $u$ over all sub-carriers, respectively.
	%
	
	On the receiver side, the received signal is sorted based on the received signal power at the given time slot to determine the decoding order. In this order, the received signal is sorted from the highest value to the lowest value. The receiver decodes and removes the UE's signal from higher received signal power to lower one. We suppose that the decoding order of UE $u$ on sub-carrier $n$ is shown by ${O}_n^{u}(k)=m$ and it means that the CU will decode the signal of user $u$ on sub-carrier $n$ after decoding the UEs that place at $\{m+1,\ldots,U\}$ order. The received signal to noise plus interference ratio (SINR) of UE $u$ on sub-carrier $n$  at time slot $k$ in the CU is denoted by $\gamma_{u}^{n}[k]$ and can be formulated as 
	\begin{align}\label{SINR}
		\gamma_{u}^{n}[k]=
		\frac{\rho_{u}^{n}[k]p_{u}^{n}[k]\big|\bold{|h\|}^2}{\sum \limits_{ O_n^{u'}(k)<O_n^{u}(k)}\rho_{u'}^n[k]p_{u'}^n[k]\big|\bold{h}_{u',n}^{H}[k]\big|^2+\sigma^2},
	\end{align}
	where  $ O_n^{u'}(k)<O_n^{u}(k)$ denotes the users that have higher decoding order to user $u$ and the users in this set can successfully adopt the SIC technique. $\sigma^2$ is the variance of zero-mean additive white Gaussian noise.
	The decoding order can be defined as 
	\begin{align}\label{SIC_con}
		\rho_{u'}^{n}[k]\big|\bold{h}_{u',n}^{H}[k]\big|^2p_{u'}^{n}[k]\leq\rho_{u'}^{n}[k]\rho_{u}^{n}[k]\big|\bold{h}_{u,n}^{H}[k]\big|^2p_{u}^{n}[k]\ .
	\end{align}
	
	By using (\ref{SINR}), the achieved data rate of  UE $ u $ at time slot $t$ is formulated as
	\begin{align}
		r_{u}[k]= \sum\limits_{n=1}^{N}\log(1+\gamma_{u}^{n}[k]). 
	\end{align}
	
	\subsection{AoI Modeling}
	In our network, in each time slot, the users send the measured information to the serving gateway, which is CU in our considered network. The data of each user is delivered to the CU via the direct link and the UAVs. Actually, the UAVs play the role of mobile relay for a source-destination pair \cite{abd2018average}.
	
	UE $ u $ generates $ d_u[k] $ (in bits) data at time slot $ k $ according to the Poisson distributions with parameter $\lambda$
	\cite{AoI_UAV_iot}. 
	Then, the UEs attempt to transmit their own data as soon as the resources are assigned.
	Data can be transmitted successfully by UE $u$ with given data rate $r_u[k]$ within the time duration $\delta$. In this case, the transmission delay of the AoI
	can be formulated as  $\frac{d_u[k]}{r_u[k]} $ in which $ \frac{d_u[k]}{r_u[k]}\le \delta $.
	Otherwise, the AoI is increased by one time slot duration.
	
	In addition, we consider that the received side can successfully decode the received signal.
	Moreover, it is supposed that the signal processing time which is needed to process the received data was neglected. If the generation time of data is defined as $k'\in\{1,2,\ldots,k\}$, the AoI evolution of UE $u$ at time slot $k$ can be expressed as:
	\begin{align}
		&\Delta_u[k]=\begin{cases}
			(k-k') \delta+\delta,& \text{if}~~\delta\ge\frac{d_u[k']}{r_u[k]}, k'\in\mathcal{K'},
			\\
			\Delta_u[k-1]+\delta, &   \text{otherwise.}
		\end{cases}
	\end{align}
	
	Note that we assume the differences between the propagation delay between relay link, UE-UAV-CU, and direct link, UE-CU, are negligible \cite{Liu1}. Also, we assume that when the UE's data can not be transmitted, the data are remains and waits until the next attempt. 
	Moreover, it is possible that some UEs do not have any data for transmission, hence, there is no need for resource allocation for that UE in that time slot.
	
	\subsection{Problem Formulation}
	In this paper, our objective is to minimize the AAoI of users over $K$ time slots by jointly optimizing the transmitting power  $\boldsymbol{P} \triangleq  \{p_u^n[k], u\in\mathcal{U}, n\in\mathcal{N},\forall k \}$, the UAVs' trajectory $\boldsymbol{Q} \triangleq \{ q_{j}[k], j \in \mathcal{J}, \forall k\} $, the reflection-coefficient elements $\boldsymbol{\Theta} \triangleq \{\beta_{l,j}[k] ,\theta_{l,j}[k], l\in\mathcal{L}, j\in \mathcal{J},\forall k \}$, and the sub-carrier assignment $\boldsymbol{\rho}\triangleq \{\rho_u^n[k],u\in\mathcal{U}, n\in\mathcal{N},\forall k \}$ subject to the mobility restrictions of the UAVs, limitations on transmit power, and constraints related to the PD-NOMA technique. The objective function can be formulated as follows: 
	\begin{align}\label{obj}
		\Psi=\max\limits_{u\in\mathcal{U}}\lim_{K\rightarrow \infty} \frac{1}{K}\sum_{ k=1}^K \Delta_u[k],
	\end{align}
	and the optimization problem can be defined as 
	\begin{subequations}\label{Main}
		\begin{align}\label{Reeq8a11}
			&\min\limits_{\boldsymbol{P},\boldsymbol{Q},\boldsymbol{\Theta},\boldsymbol{\rho}}\;~\Psi, 
			\\
			\text{s.t.}~~& \text{C1, C2, C3, C4, C5, C6,}\nonumber\\
			\label{SIC_Con1}\text{C7:~}& 	\rho_{u'}^{n}[k] \big|\bold{h}_{(u',n)}^{H}[k]\big|^2p_{u'}^{n}[k]\le \rho_{u'}^{n}[k] \rho_{u}^{n}[k]\big|\bold{h}_{(u,n)}^{H}[k]\big|^2p_{u}^{n}[k]\nonumber\\
			&~~~~~~~~~~~~~ \text{~~if~~} O_n^{u'}(k)<O_n^{u}(k), \forall u',u, n,k, \\
			\label{theta_1}\text{C8:~}& 0\le \theta_{1,j}\le 2\pi, ~~~~~~~~~~~~\forall j,\\
			\label{beta_1}\text{C9:~}& \beta_{l,j} \in(0,1] ~~~~~~~~~~~~~~~~\forall j,\\
			\label{am}\text{C10:~}&	\rho_{u}^{n}[k] \in\{0,1\}, ~~~~~~~~~~~~\forall n,j,
		\end{align}
	\end{subequations}
	where 
	\eqref{SIC_Con1} is SIC-related constraint.
	The range of phase shift and amplitude of the reflecting elements are defined in \eqref{theta_1} and \eqref{beta_1}. Finally, \eqref{am} defines the sub-carrier allocation variable as a distinct variable.
	
	
	

	\section{Solution Method}
	\label{Solution}
	Problem \eqref{Main} is a non-convex optimization problem and incorporates both integer and continuous variables which is difficult to solve. Therefore, finding a low complex and efficient solution
	is complicated. To solve \eqref{Main}, we propose a  deep reinforcement learning (DRL) that has shown high potential in solving a variety of complex problems in the context of UAV communications. To apply the DRL-based solution, we transform the problem \eqref{Main} in the form of MDP, then devise a DRL-based technique to solve it.
	
	Firstly, we explain the elements of the RL, and the MDP, in the following \cite{RL_Book, IRS_AI}. Generally, the environment and the agent are two main parts of the RL. The agent during numerous time slots/epochs learns environment.
	We suppose that the CU is the agent that performs the actions in a central manner.



	We define the states, actions, and reward function of the problem \eqref{Main} as follows:
	\\\textbf{States:}  The state of the environment $s_k\in \mathcal{S}$ at the beginning of the $k$-th time slot is defined as
	\begin{align}
		{s}_k=\{\boldsymbol{q}[k-1],\boldsymbol{d}[k], \boldsymbol{\Delta}[k]\} , 
	\end{align}
	where $\boldsymbol{q}[k-1] \triangleq [q_1[k-1],\ldots,q_J[k-1]]^T$, $\boldsymbol{d}[k] \triangleq[d_1[k],\ldots,d_U[k]]^T$, and $\boldsymbol{\Delta}[k] \triangleq[\Delta_1[k],\ldots,\Delta_U[k]]^T$. The state includes the current position of the UAVs, 
	the amount of data of all users, the value of the AAoI related to all users.
	\\\textbf{Actions:} Actions are the optimization variables, i.e. the position of UAVs, the configuration of IRS (phase shift and angle), power allocation, and sub-carrier assignment. Let $a_k=\{\bold{q}[k],\boldsymbol{\theta}[k], \bold{p}[k],\boldsymbol{\rho}[k]\}$ be the  actions at  time slot $k$. Let us define $\boldsymbol{\theta}[k]\triangleq \{\beta_{l,j}[k] ,\theta_{l,j}[k], l\in\mathcal{L}, j\in \mathcal{J}\}$, $\bold{p}[k]\triangleq \{p_u^n[k], u\in\mathcal{U}, n\in\mathcal{N} \}$, and $\boldsymbol{\rho}[k] \triangleq \{ q_{j}[k], j \in \mathcal{J}\}$.
	The action includes both integer, i.e. $\boldsymbol{\rho}[k]$, and continuous variables, i.e. $\boldsymbol{\theta}[k]$ and $ \bold{p}[k]$. We note that the action space includes all valid actions which fulfill the constraints of \eqref{Main}.
	\\\textbf{Reward Function:}   In general,  the reward function is a numerical value obtained by the agent from the environment and amount the impact of current action. Since the main purpose is to minimize the AAoI, we consider the value of \eqref{obj} as the reward. 
	The maximum freshness (as negative of  AAoI) is defined as reward function as follows:
	\begin{align}\label{Reward}
		R_k = -\max\limits_{u\in\mathcal{U}}\sum_{ k=1}^K \Delta_u[k].
	\end{align}
	
	The reward is calculated based on  ${a}_k$ which is decided by observing state  $s_k$, then the environment transits to the next state, $s_{k+1}$. It is worthwhile to mention that the state transition is a random process, with transition probability denoted by $\Pr\big(s_{k+1}|s_k,a_k\big)$. The randomness in our model comes from the probability of taking actions and the randomness related to the network, i.e, changing the traffic of users as well as their channel gains.  
	\subsection{MDP and Policy Algorithm}
	Generally, MDP describes an environment that is perfectly observable, i.e., the observed state can completely characterize the features of the environment. The main aspect of the MDP is that the action is chosen based on the current state and the history of states has no effect on decisions \cite{Cooperative}.
	
	The goal in the solution of the MDP is to find the optimal policy that teaches the agent how to maximize the reward function by choosing the next action based on the current state. Basically, the policy is a mapping between states to the actions that are determined by a probability distribution over the action space. Considering the action, state, and the reward function, let $\pi(s_k), s_k\in \mathcal{S}$ denotes the policy function.
	
	Due to the fact that the transition probability is unspecified, we adopt the Q-learning method. 
	Which is a subset of the RL methods.
	The Q-function is defined as the value to taking action $a_{k}$ in the state $s_k$ under given policy $\pi(s_k)$ and indicates the expected cumulative reward after taking action $a_k$ under state $s_k$ as follows:
	\begin{align}
		Q^{\pi}(s_k, a_k)=\mathbb{E}\Big[ \sum\limits_{i=0}^{\infty} \gamma^{i}R_{k+i} | s_k=s, a_k=a\Big].
	\end{align}
	
	By using the Bellman equation, which can be formulated as 
	\begin{align}
		Q^{\pi}(s_k, a_k)=\mathbb{E}\big[R_k+\gamma Q^{\pi}(s_{k+1}, a_{k+1})\big].
	\end{align}
	
	Let's denote $Q^{\star}(s, a)$ as the optimal $Q$-function by following $\pi^{\star}$ where $Q^{\star}(s_k, a_k)=Q^{\pi^\star}(s_k, a_k) = \max_{\pi}Q^{\pi}(s_k, a_k) $.
	
	
	Thanks to recent achievements on the neural networks, deep neural networks (DNNs) are introduced to mitigate the issue of large action state spaces and can be used as function approximators to predict the optimal Q-function as $Q^{\star}(s_k,a_k)\approx Q(s_k,a_k;\boldsymbol{w})$, where  $\boldsymbol{w}$ is the vector of the weights of the neural (deep neural)  \cite{DQNMAIN}.
	While DQN works well on discrete action spaces, for continuous there are some reductions in the performance \cite{DQNMAIN}. 
	Meanwhile, as mentioned before, our action space includes both discrete and continuous. Therefore, we have challenges in finding a DRL-based method that handles both discrete and continuous, and scales well with the numbers of users.
	In the following subsection, we will discuss our devised DRL-based algorithm.
	\subsection{Proposed DRL-based Algorithm }
	As discussed before, our original problem incorporates both integer and continuous high dimensional variables. Actually, the DRL methods do not have appropriate functionality in a problem with mixed-integer and continuous variables. 
	
	Inspired by \cite{Cooperative}, we propose the neural combinatorial method where discrete action $a_k^{\text{Dis}}$ and continuous actions $a_k^\text{Con}$ are derived separately. Actually, this concept is based on the well-known iterative optimization which has low complexity and can deal with more complicated problems. The rationale behind this method is dividing the main problem into two sub-problems, then optimizing each of them with corresponding variables.
	
	At first, we divide the action space into two action spaces, discrete and continuous. Then, we perform DRL-based optimization to find the corresponding actions,  applied distinct
	methods DRL-based for each separate actions have a distinct method  DRL-based.
	Thanks to the advanced results on the RL \cite{Cooperative,DDQN1}, we propose a double DQN  (DDQN) method for our discrete action space and a proximal policy optimization (PPO) for the continuous action space \cite{PPO-Basic,PPO-RIS-AoI}. 
	
	$\bullet~$\textbf{DDQN for Discrete Actions:}
	In DDQN, there are two neural networks (NNs) named as main and target networks with weights $\boldsymbol{w}$ and $\boldsymbol{w}^{\prime}$, respectively. The function of the main network approximates the Q-function and performs the action selection. The target network stabilizes the performance of the learning network. At each time slot $k$, the agent chooses action $a^{\text{Dis}}_{k}$ based on state $s^{\text{Dis}}_k$ and perceives the reward $R^{\text{}}_{k}$ and new state $s^{\text{Dis}}_{k+1}$ which represent changes taking place in the environment after taking the action. The set $(s^{\text{Dis}}_k,a^{\text{Dis}}_k,R^{\text{}}_k,s^{\text{Dis}}_{k+1})$ which is known as an experience sample is stored in an experience memory related to the DDQN algorithm with capacity $D_{\text{DDQN}}$. When the number of samples reaches the adequate amount, $D_s$ samples (mini-batch) are selected randomly to minimize the loss function and update the weight of main network. The loss function is defined as \cite{DDQNfrist}
	\begin{align}\label{loss_function_DDQN}
		L(\boldsymbol{w})^{\text{DDQN}} = \mathbb{E} \Big[ \big(y^{\text{DDQN}} - Q(s^{\text{Dis}}_k,a^{\text{Dis}}_k;\boldsymbol{w}) \big)^2 \Big],
	\end{align}
	where 
	\begin{align}
		y^{\text{DDQN}} = R^{\text{}}_{k}+\gamma Q\left(s^{\text{Dis}}_{k+1}, \underset{a^{\text{Dis}}_{k+1}}{\operatorname{argmax}} Q\left(s^{\text{Dis}}_{k+1}, a^{\text{Dis}}_{k+1} ; \boldsymbol{w}\right), \boldsymbol{w}^{\prime}\right)^2.
	\end{align}
	
	%
	In our case, the output of DDQN is the discrete action when the continuous actions have been known (obtained via PPO). The discrete action selection is done via taking the maximization operator over the estimated Q-function. 
	
	$\bullet~$\textbf{PPO for Continuous Actions:}
	Recently focus on the RL has shown that the PPO is a powerful tool to solve continuous actions. The PPO uses a new objective function and the implementation and tuning tasks are more convenient than other policy gradient methods \cite{PPO-Basic, PPO2}.   
	These are a few works that applied PPO in the optimization, for example, the  AoI optimization in aerial RIS via the PPO are studied in  \cite{PPO-RIS-AoI}. 
	These results motivate us to customize the PPO for our continuous actions and analyze its performance in the context of the wireless networks. 
	In the PPO, the learning process includes iteration phases in which each iteration includes $K_{\max}$ time slots. In each learning iteration, the PPO agent alternates between the sampling and optimizing phases over $K_{\max}$ time slots and performs optimization on its own action and neural networks parameter $\boldsymbol{\varpi}$. 
	During each time slot, after observing the current state of the network and choosing the appropriate action, an advantage estimate is computed to improve the training efficiency of the policy \cite{PPO-Basic}.
	
	To update the policy $\pi_{\boldsymbol{\varpi}}$, the objective function defined by \eqref{PPOObj} (at top of next page) is computed, 
	\begin{figure*}
		\begin{align}
			\label{PPOObj}
			L^\text{CLIP}(\boldsymbol{\varpi})=\mathbb{E}_{k}\left[\min \left(\frac{\pi_{\boldsymbol{\varpi}}(a^{\text{Con}}_k \mid s^{\text{Con}}_k)}{\pi_{\boldsymbol{\varpi}_\text{old}}(a^{\text{Con}}_k | s^{\text{Con}}_k)} \mathfrak{A}_{\pi_{\boldsymbol{\varpi}_\text{old}}}(s^{\text{Con}}_k, a^{\text{Con}}_k),\right.\right.
			\left.\left.\operatorname{clip}\left(\frac{\pi_{\boldsymbol{\varpi}}(a^{\text{Con}}_k \mid s^{\text{Con}}_k)}{\pi_{\boldsymbol{\varpi}_\text{old}}(a^{\text{Con}}_k |s^{\text{Con}}_k)}, 1+\epsilon, 1-\epsilon\right) \mathfrak{A}_{\pi_{\boldsymbol{\varpi}_\text{old}}}(s^{\text{Con}}_k, a^{\text{Con}}_k)\right)\right],
		\end{align}
		\hrule
	\end{figure*}
	where $\epsilon$ is the clip fraction used to control the clip range and $\mathfrak{A}(s^{\text{Con}}_k, a^{\text{Con}}_k)$ is the estimated advantage function in episode $k$. Moreover, the term $\operatorname{clip}\left(\frac{\pi_{\boldsymbol{\varpi}}(a^{\text{Con}}_k \mid s^{\text{Con}}_k)}{\pi_{\boldsymbol{\varpi}_\text{old}}(a^{\text{Con}}_k \mid s^{\text{Con}}_k)}, 1+\epsilon, 1-\epsilon\right)$ modifies the objective by clipping the probability ratio, which limits the variation of policy into the interval $[1−\epsilon,1 + \epsilon]$. For more details about the structure and mechanism of PPO agent, please refer to the \cite{PPO-Basic}.
	
\begin{figure*}[!ht]
		\centering
		{\includegraphics[width=.90\textwidth]{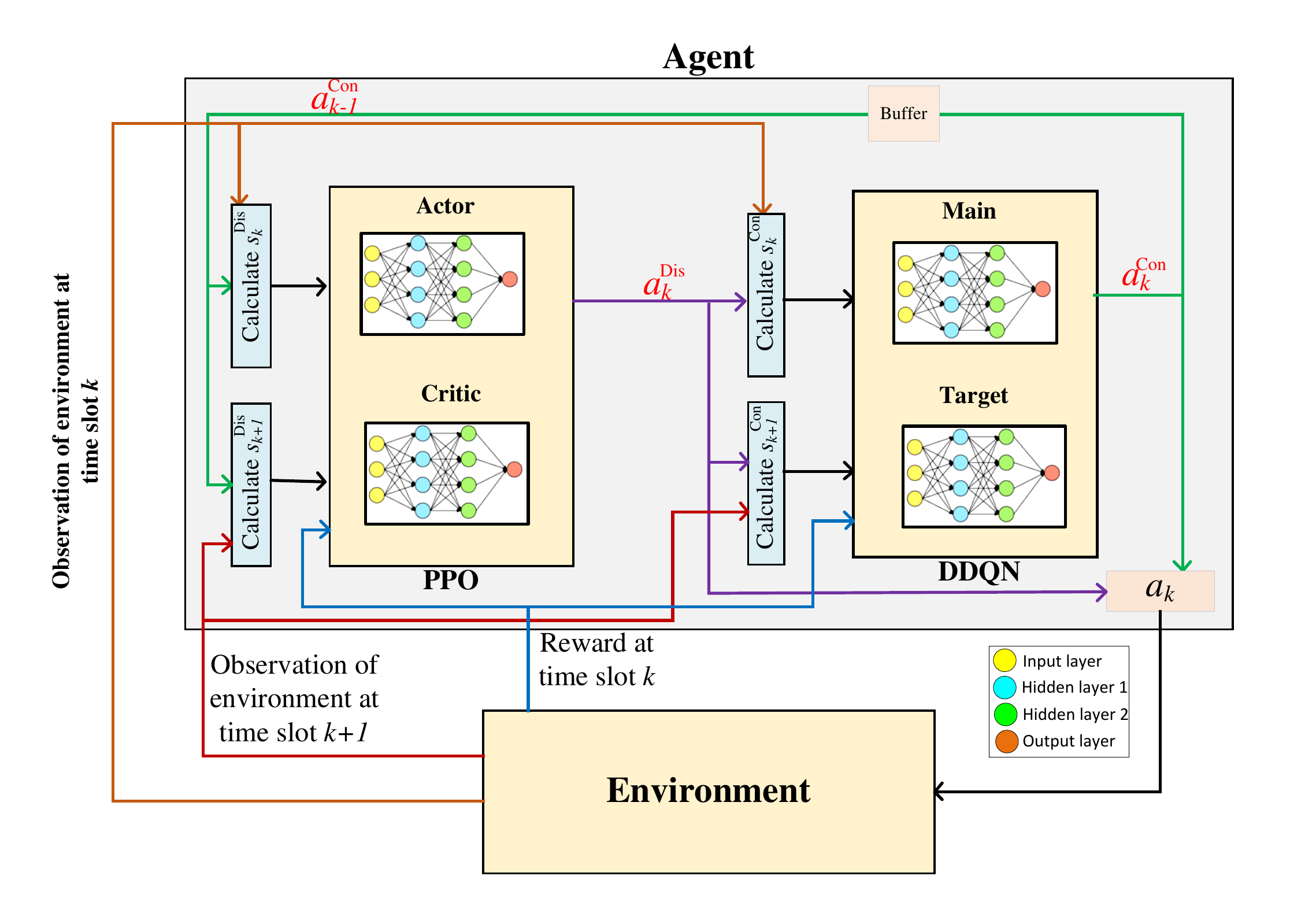}}
		\caption{The schematic illustration of the proposed DRL-based solution.}
		\label{flochart}
	\end{figure*}

	
	The main steps of the solution are illustrated in Fig. \eqref{flochart} and Algorithm \ref{DDQN_PPO_algorithm}.

	\begin{algorithm}
		\renewcommand{\arraystretch}{0.9}
		\caption{DDQN-PPO mixed algorithm}
		\label{DDQN_PPO_algorithm}
		Initialize the experience memory for DDQL and PPO with capacity $D_{\text{DDQL}}$ and $D_{\text{PPO}}$\;
		Initialize $\boldsymbol{w}$ and $\boldsymbol{\varpi}$ randomly and set $\boldsymbol{w}^\prime\leftarrow\boldsymbol{w}$ and $\boldsymbol{\varpi}_{old}\leftarrow\boldsymbol{\varpi}$\;
		Set $D=1$ as number of experience in experience memories\;
		\For {$k=1:K_{\max}$}
		{
			Calculate current network state $s^\text{Dis}_k$ based on $a_{k-1}^{\text{Con}}$\;
			Based on $\epsilon$-greedy select action $a^\text{Dis}_{k}$\;
			Observe next state $s^\text{Dis}_{k+1}$\;
			Set $s^\text{Con}_{k}=s^\text{Dis}_{k+1}$\;
			Sample action $a^{\text{Con}}_k$ via policy $\pi_{\boldsymbol{\varpi}_{}}$\;
			Get relevant reward $R^{\text{}}_k$ and $s^{\text{Con}}_{k+1}$ based on $a^\text{Dis}_{k}$ and $a^\text{Con}_{k}$\;
			Store experience $(s^\text{Dis}_k,a^\text{Dis}_k,R^\text{}_k,s^\text{Dis}_{k+1})$ and  $(s^{\text{Con}}_{k}, a^{\text{Con}}_k, R^{\text{}}_k, s^{\text{Con}}_{k+1})$  in DDQL and  PPO memories, respectively\;
			Set $D\leftarrow D+1$\;
			\If {$D>{D_s}$}{
				Sample a mini-batch from DDQL memory experiences ramdomly\;
				Calculate \eqref{loss_function_DDQN} and minimize it to update $w$\;
				Sample a mini-batch from PPO memory experiences ramdomly\;
				Compute and optimize  PPO's objective function given \eqref{PPOObj}\;
				Update weight of target network $w^\prime$ by\;
				$w'=\tau w +(1-\tau)w^\prime$ every predefined time slot\; 
				Update $\boldsymbol{\varpi}_{old}$ by every predefined time slot:\;
				$\boldsymbol{\varpi}_{old} \leftarrow \tau\boldsymbol{\varpi}+(1-\tau)\boldsymbol{\varpi}$\;
			}
		}
	\end{algorithm}

	\subsection{Complexity and Convergence }
	The complexity analysis is described in this subsection.
	In practice, the PPO frequently is implemented in an Actor-Critic framework. More objective functions are supplied to the surrogate goal.
	The amount of multiplications in each iteration is commonly used to calculate the complexity of DRL techniques like the Actor-Critic framework.  
	We suppose that a learning network consists of $L_{\text{Hid}}$ hidden layers where the neurons of hidden layer $l$ can be denoted by $n_l$. Actually, the number of operations at each iteration determines the complexity of learning networks. The numbers of neurons are denoted by $n_l$, $l=1,\dots,{L_{\text{Hid}}}$. 
	The complexity of each machine learning network is $\mathcal{O}\Big(\sum\limits_{l=1}^{L_{\text{Hid}}-1}n_l n_{l+1}$\Big).\\
	The convergence of learning algorithms highly depends on the hyperparameters in the training process. Hence, finding a theoretical analysis is hard. The convergence of our learning algorithm is shown via numerical simulation results in Fig. \ref{Rewared.}.
	
	\section{SIMULATION AND NUMERICAL ANALYSIS}
	\label{Simulation_Results}
	In this section, we numerically evaluate our proposed algorithm in the studied framework. \footnote{The implementation of simulation is provided on:
http://ieee-dataport.org/8593
}
	For the simulation setup, we consider a square area of  $400$ m $\times$ $400$ m, where $U=20$ users are randomly distributed with  uniform distribution. The initial location of UAVs is selected randomly inside the allowed area. The UAVs fly at height $50$ m above the ground. 
	Moreover, we consider that the CU is located at point $(0,0,10 \textrm{m})$ \footnote{Actually the CU with 10m antenna height is located at point (0,0,10). Hence, it is obvious that we can consider that the location of the CU is  $(0,0,10 \textrm{ m})$.}. The minimum distance between two UAVs is set to $ D_{\min}=8$ m. The maximum speed of each UAV is  $10$ m/s and they only can serve within the area with a distance of $400$ m from CU. 
	
	The number of orthogonal sub-carriers is $N=4$, and the number of antennas $M=2$. We assume all users have the same maximum transmit power which is configured to $20$ dBm. Other parameters are listed in Table \ref{tab:plan}. Results are obtained by using the PyTorch library, and the parameters of the learning algorithm are listed in Table \ref{Ta_PPo_Pa}. 

	\begin{table}
		\caption{Simulation Parameters}
		\label{tab:plan}
		\centering
		\begin{tabular}{|p{4.2cm}|p{2cm}|}
			\hline
			\textbf{Parameter} & \textbf{Value}  \\
			\hline
			The number of reflector elements ($L$) & 100  \\
			\hline
			$ d/ \lambda$ & 0.5\\
			\hline
			$z_{u}^n$ & $10{e}^-15$ dBm   \\
			\hline
			$ \delta$ & 100 s  \\
			\hline
			$ P_{\text{mask}}^{n}$ & 5 dBm  \\
			\hline
			
			
			Band-width & 200 KHz \\
			\hline	
			$ \kappa_n$ & 3  \\ 
			\hline
		\end{tabular}
	\end{table}
	\begin{table}
		\caption{Learning parameters}
		\label{Ta_PPo_Pa}
		\centering
		\begin{tabular}{|p{4cm}|p{2cm}|}
			\hline
			\textbf{Parameter} & \textbf{Value}  \\
			\hline
			Reward discount ($\gamma$) & 0.8   \\
			\hline
			Layer1-size & 400   \\
			\hline
			Layer2-size  & 30   \\
			\hline
			Episode & 4000 \\
			\hline
			Number of time slot & 600\\
			\hline
			Learning rate for actor & 0.00001  \\ 
			\hline
			Learning rate for critic& 0.0001 \\ 
			\hline
			Soft-replacement$(\tau)$ & 0.01      \\ 
			\hline
			Memory size of DDQN and PPO & 10000    \\
			\hline
			Batch size of DDQN and PPO & 128 \\
			\hline
			
		\end{tabular}
	\end{table}

\begin{figure}[!ht]
		\centering
		{\includegraphics[width=0.48\textwidth]{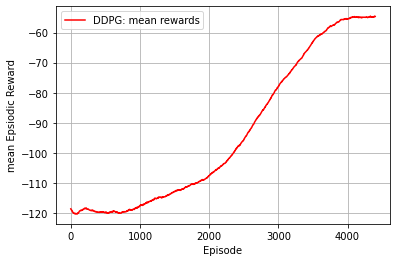}}
		\caption{An illustration of the convergence behavior of the proposed  algorithm.}
		\label{Rewared.}
	\end{figure}

	First, we verify the convergence behavior of the proposed algorithm shown in Fig. \ref{Rewared.}. The negative reward is because of considering AoI in the reward function. It can be seen that the adapted agent improves the reward as the learning time grows, and convergence to a stable point with acceptable speed (i.e., the learning time is not much) is achieved. Thus, our algorithm can deal with environments with no exact knowledge of its dynamic, where in some cases obtaining exact knowledge of the underlying dynamic is quite complex. 

\begin{figure}[!ht]
		\centering
		{\includegraphics[width=0.48\textwidth]{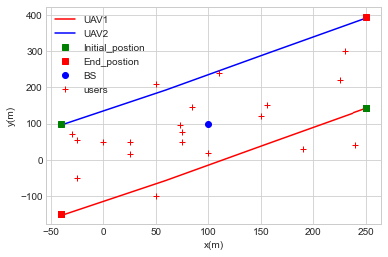}}
		\caption{An example of trajectory design of UAVs.}
		\label{trajectory.}
	\end{figure}

\begin{figure}[!ht]
		\centering
		{\includegraphics[width=0.48\textwidth]{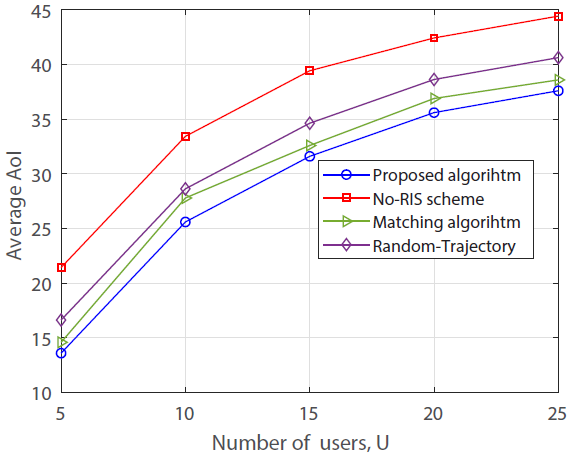}}
		\caption{AoI versus the number of total users.}
		\label{Impact_Users}
	\end{figure}

\begin{figure}[!ht]
		\centering
		{\includegraphics[width=0.48\textwidth]{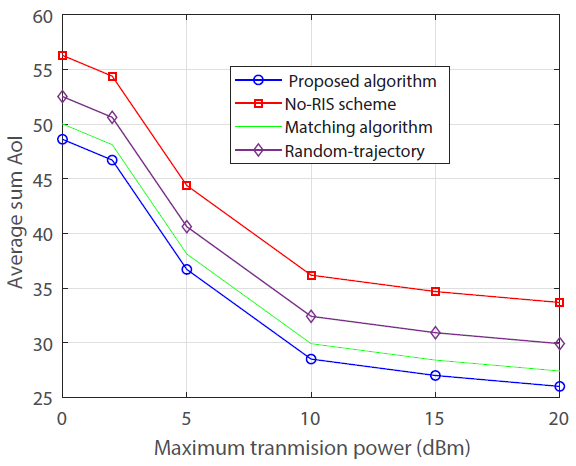}}
		\caption{Sum of AAoI versus the maximum transmit power (dBm).}
		\label{Impact_Power}
	\end{figure}

	In Fig. \ref{trajectory.}, we illustrate an exemplary trajectory, denoted by solid lines, of UAVs in the  coverage area with the given user's locations.  It is observed that optimizing trajectory is in the direction of making UAVs near to users that are far from the CU, to improve the wireless channel, resulting in the improving date rate and the AAoI reduction.
	\\\indent
	To compare our algorithm and considered system model, we consider the following baselines: 1) \textit{No-RIS scheme}, where we consider standalone UAV without RIS elements \cite{AoI-ARIS-1},\cite{PPO-RIS-AoI} 
	2) \textit{Random-trajectory}, where the trajectory of UAV is determined randomly regardless of the current state; which is similar to random walk policy in \cite{PPO-RIS-AoI},
	3) \textit{Matching} algorithm, where sub-carriers are assigned to users according to the matching algorithm \cite{Matching}. Moreover, we schedule users with highest AoI, i.e., greedy policy, so that the overall size of status data does not exceed the throughput capacity.
	\\\indent
	We plot the AAoI with respect to the number of total users in Fig. \ref{Impact_Users} for different algorithms. First, we can observe that the AAoI grow as the number of users increase, due to the limited wireless capacity of the considered network under given available resources, i.e., only a limited number of users can be scheduled to send their own status data at each time. Consequently, the waiting time to send new updates is increased generally; hence, the AAoI is also increased.  In this figure, we observe that the proposed algorithm outperforms other baseline algorithms, e.g., the No-RIS scheme and random trajectory. This is because, in our scheme, we both exploit the RIS technology and optimize trajectory, gaining an increased throughput of the network; hence, it reduces the waiting time of update for all users at all.  However, increasing throughput does not necessarily improve the AAoI; because it is also important to optimize the users who are scheduled, where we use a greedy-based algorithm.
	\\\indent
	Fig. \ref{Impact_Power} plots the sum of AAoI versus the maximum transmission power per user for different algorithms. This figure shows the impact of the amount of transmit power on the AAoI reduction, for some specific ranges of the power. The reason is that by increasing the amount of transmit power, the SINR value will be increased, hence the throughput of the network improves and the AAoI reduces.  
	On the other hand, increasing power leads to more interference in the network which reduces wireless capacity. An important result from 	Fig. \ref{Impact_Power}, when the power budget is small, employing the RIS is important, where in many applications such as power-limited sensor networks power is a pivotal resource it is not allowable to use much powers. It means in such cases, we can use some alternative advanced technologies, such as RIS to enhance network performance from the AAoI perspective. 
	
	\section{Conclusion}
	\label{CONCLUSION}
	In this paper, we studied a status update framework for IoT users in an aerial RIS communication systems. To this end, we focused on generated updates of deliver users randomly to the intended destination timely, with help of aerial RIS-embedded stations, where they relay information to the destination over wireless channels. In this regard, we formulated an optimization problem aiming to minimize the AAoI by optimizing trajectory, sub-carrier, and power variables under system-wide and budget constraints. As a solution for our mixed-integer non-convex problem, we adopted a state-of-the-art learning method based on the policy optimization called PPO-agent. Numerical analysis was provided to investigate our algorithm and compare its performance with other baseline algorithms. Simulations results showed the effectiveness of our algorithm and system compared to baselines algorithms.
	
	\bibliographystyle{IEEEtran}
	\bibliography{Bibliography}
	\clearpage
	‏
\end{document}